\documentclass[epj-spec]{svjour}
\usepackage{graphics}
\begin{document}
\title{Dark Halo or Bigravity?}
\author{Nicola Rossi \thanks{\email{nicola.rossi@aquila.infn.it}}}
\institute{Dipartimento di Fisica, Universtit\`a degli Studi dell'Aquila, 
I-67010 Coppito (AQ), Italy \\ 
and INFN, Laboratori Nazionali del Gran Sasso, I-67010 Assergi (AQ), Italy}
\abstract{
Observations show that about the $20\%$ of the Universe is
composed by invisible (dark) matter (DM), for which many
candidates have been proposed. In particular, the anomalous
behavior of rotational curves of galaxies (i.e. the
flattening at large distance instead of the Keplerian fall)
requires that
this matter is distributed in an extended halo around the galaxy. 
In order to reproduce this matter density profiles in Newtonian gravity and 
in cold dark matter (CDM) paradigm (in which the
DM particles are collisionless), many ad-hoc 
approximations are required.
The flattening of rotational curves can be explained by a
suitable modification of gravitational force in bigravity theories,
together with mirror matter model that predicts the existence
of a dark sector in which DM has the same physical 
properties of visible matter.
As an additional result, 
the Newton constant is different at distances much less and 
much greater than 20 kpc. 
[\emph{To be published in Proc. Int. Conf. "Atomic Clocks and Fundamental 
Constants" ACFC 2007 - Bad Honnef, Germany, June 2007}]
} 
%
\maketitle
\section{Introduction}

Cosmological observations show the Universe to be nearly flat,
i.e. the energy density is very close to the critical one: 
$\Omega_{tot} \simeq 1$. We can separate the different contributions
present in the total density. Mainly, we can write 
$ \Omega_{tot} = \Omega_M + \Omega_{\Lambda} $, 
in which we distinguish the contributions due to
matter ($\Omega_M=0.24\pm 0.02$) and to the dark energy 
(cosmological term, $\Omega_\Lambda=0.76\pm 0.02$). In 
particular $\Omega_M=\Omega_B+\Omega_D$, where 
$\Omega_B=0.042\pm 0.005$ and 
$\Omega_D=0.20\pm 0.02$ are, respectively, the baryonic
and dark matter components \cite{PDG}.

Let us focus on the dark matter problem. 
CMD model is in agreement with  
experimental results on Cosmic Microwave 
Background (CMB) and
Large Scale Structure (LSS) that confirm the presence of the 
dark matter component.
Many DM candidates have been proposed in the literature as 
axions ($m\sim 10^{-5}$ eV), neutralinos ($m\sim$ TeV), 
wimpzillas ($m\sim 10^{14}$ GeV) and so on. 
However,   the question why the fractions of the 
baryonic and dark components are so
close, $\Omega_D/\Omega_B\sim 5$, remains unresolved.  
  
A most convincing proof for DM comes from  
rotational curves of galaxies and cluster dynamics.
In the Newtonian picture, at large distance from the center
of a galaxy one expects that the velocity 
along a circular orbits behaves as $ v(r) \propto 1/\sqrt{r} $
(also known as Keplerian fall). 
This is due to the fact that the gravitational potential of
a galaxy outside the inner core (bulge) is 
$\phi \simeq GM/r$, where $G$ is the Newton constant. 
Instead, one observes that in these regions $v$ is
approximately constant. In order to explain this anomalous
behavior, without modifying the Newtonian paradigm,
one has to suppose the existence of dark matter
distributed in a extended spherically symmetric halo around the galaxy,
according to the ``isothermal'' mass distribution 
profile $\rho(r) \propto (1+(r/a)^2)^{-1}$,
where $a$ is a scale radius. Since gravity is universal
between visible and dark matter, a point-like source of both 
types of matter generates a potential 
\begin{equation} \label{eq:cpot}
  \phi(r) = \frac{G}{r}(\mathcal{M}_1+\mathcal{M}_2)
\end{equation}
where $\mathcal{M}_1$ and $\mathcal{M}_2$ are, respectively, the 
visible and dark components.
CDM model assumes that
the DM particles are collisionless. Due to this property,
N-body numerical simulations tend actually to
predict a different behavior: the DM distribution has
a cusp profile $\rho(r) \propto 1/r^\alpha$, with $\alpha=1\div1.5$
\cite{nfw,sal1}. For many galaxies these cusp profiles do not reproduce
the observed rotational curves as well as the isothermal profile.

An alternative possible candidate is the Mirror matter \cite{mirror}. 
According to this model 
our Universe is made of two similar gauge sectors. 
In other words,  
in parallel to our sector of the ordinary particles (O-) and interactions 
described by the Standard Model, there exists a hidden sector (M-) 
that is an  exact duplicate of ordinary sector
in which particles and interactions have exactly the same characteristics, 
and the two sectors are connected by the common gravity  
(see for review \cite{alice}). 
Therefore, if the mirror sector exists, then the Universe 
should contain along with the ordinary particles (electrons, nucleons, 
photons, etc.) also their mirror partners with exactly the same masses 
(mirror electrons, mirror nucleons,  mirror photons, etc.). 
Mirror matter, invisible in terms of ordinary photons, can naturally 
constitute dark matter. 
One should stress that 
the fact that O- and M-sectors have the same microphysics, 
does not imply that their cosmological evolutions should be the same too.
In fact, if mirror particles had the same temperature  
in the early universe as ordinary ones, this would be in immediate 
conflict  with Big Bang Nucleosynthesis (BBN). 
The BBN limit on the  effective number of extra neutrinos 
implies that the temperature of the mirror sector $T'$ must be at least 
about 2 times smaller than the temperature $T$ of the ordinary sector,  
which makes mirror baryons viable candidate for dark matter. 
In particular, the mirror dark matter scenario would 
give the same pattern of LSS and  
CMB as the standatd CDM 
if $T'/T < 0.2$ or so \cite{cosmology}. 
In addition, the baryon asymmetry of the Universe can be generated 
via out-of-equilibrium $B-L$ and $CP$ violating processes between 
ordinary and mirror particles \cite{bariogenesi} 
whose mechanism could explain 
the intriguing puzzle of the correspondence between 
the visible and dark matter fractions in the Universe, 
naturally predicting the ratio
$\Omega_D/\Omega_B\sim 1 \div 10$  \cite{fractions}. 

However, in contrast to the collisionless CDM, mirror baryons 
obviously constitute collisional and dissipative dark matter. 
Therefore, one should expect that mirror matter undergoes 
a dissipative collapse 
and thus in the galaxies it is distributed in a similar manner 
as the visible matter 
instead of producing extended quasi-spherical CDM halos. 
Indeed even the hidden sector undergoes a dissipative
collapse as the visible sector that follows
an exponential profile $\rho(r) \propto e^{-r/r_0}$.
In this way the distribution of the dark matter is more compact 
in the center of the galaxy and is not extended as the 
CDM halo.
 
Since gravity is universal between the two
sectors, this mirror dark matter hypothesis gets into difficulties 
to explain the flat rotational curves of galaxies. 
However we can 
suppose that each sector has its own gravity and that
mixing term produces a suitable modification of
gravity at large distance
\footnote{The particle mixing phenomena between ordinary and 
mirror sectors were discussed in the literature for photons 
\cite{photons}, neutrinos \cite{neutrinos}, neutrons \cite{neutrons}, 
etc., as well as possible common gauge interactions between 
two sectors \cite{gauge}. The 
mixing between the ordinary and mirror gravitons was 
first discussed in our recent papers \cite{us01,us02}.}.  
In particular we show that 
the interaction term allows us to obtain a massive graviton and
leads to a modified potential. A test mass of type 1 at distance
$r$ from the origin in which there is a sources of both  
types of matter ($\mathcal{M}_1$ and $\mathcal{M}_2$),
instead of (\ref{eq:cpot}),  
feels a potential
\begin{equation} \label{bigr}
  \phi(r)= \frac{G}{2r} 
  \left( \mathcal{M}_1 + \mathcal{M}_2  \right) +
  \frac{G e^{-\frac{r}{r_m}}}{2r} 
  \left( \mathcal{M}_1 - \mathcal{M}_2 \right),
\end{equation}
where $G$ is the Newton constant and $r_m$ is the range of
the massive graviton. Notice that at small distance 
$r \ll r_m$ the test mass interacts only with $\mathcal{M}_1$
through the ordinary Newton potential, whereas, at large distance 
$r \gg r_m$ the test mass interact with the sum of the two kinds
of matter $(\mathcal{M}_1 + \mathcal{M}_2)/2$. This result,
together with the mirror matter hypothesis, enables us to
reproduce the observed rotational curves of galaxies.  

\section{The Model} 

Let us consider a theory with two dynamical metrics $g_{1,2 \, \mu\nu}$,
each of them interacting with its own matter. The total action
contains two Hilbert-Einstein actions and a mixed term $\mathcal{V}$:  
\begin{equation} \label{action}
  \mathcal{S}=\int d^4x \left[ \sqrt{g_1}(\frac{M_1^2}{2} R_1 +
  \mathcal{L}_1)+ \sqrt{g_1}(\frac{M_2^2}{2} R_2 + 
  \mathcal{L}_2) -\mu^4 (g_1 g_2)^{1/4} \mathcal{V}(g_1,g_2) \right],
\end{equation}
where $M_{1,2}$ are the Planck masses (in general different) and
$\mathcal{L}_{1,2}$ are the corresponding matter Lagrangians 
(respectively, ordinary matter or type 1  and dark matter or type 2).
The action (\ref{bigr}) describes a more generic bigravity theory,
that is the simplest case of the multigravity theory which 
considers $ N $ metrics interacting each other through a mixing
term (see for review \cite{dam_kog}). 
The interaction term breaks down the invariance under the diffeomorphism
group $D_1 \otimes D_2$ to a diagonal diffeomorphism 
$D_{1+2}$.
The two metrics $g_1$ and $g_2$, in the flat (Minkowski) space 
approximation, can 
be written as $g_{1,2\mu\nu} \simeq \eta_{1,2\mu\nu} 
+ h_{1,2\mu\nu} / M_{1,2}$. 
The mixed term in (\ref{action}) can induce the non diagonal rank 1 mass matrix
between two gravitons $h_1$ and $h_2$ which has one massless and 
one massive eigenstates:   
\begin{equation} \label{eq:rotation}
   \left\{\begin{array}{ccc}
    h_{\mu\nu} & = & \cos\vartheta h_{1\mu\nu} + \sin\vartheta h_{2\mu\nu} \\
    \tilde{h}_{\mu\nu} & = & 
    -\sin\vartheta h_{1\mu\nu}+\cos\vartheta h_{2\mu\nu} 
   \end{array} \right.
\end{equation} 
where $\vartheta$ is a mixing angle: $\tan \vartheta = M_2/M_1$. 
In the case $M_1=M_2$, i.e. $\vartheta=\pi/4$,
the rotation (\ref{eq:rotation}) reduces
to even and odd combinations of $h_{1,2\, \mu\nu}$.
The massless state $h_{\mu\nu}$ is the ordinary graviton
that exhibits a Newtonian potential $\sim 1/r$
universally coupled with both matters.

The massive state $\tilde{h}_{\mu\nu}$ in turn can have a Lorentz 
breaking (LB) mass pattern \cite{rubakov} 
\begin{equation} \label{rub}
  \mathcal{L}_{\rm mass} =\frac{M_{Pl}^2}{2} \left( m_0^2 \tilde{h}_{00}^2 + 
  2 m_1^2 \tilde{h}_{0i}^2 -
  m_2^2 \tilde{h}_{ij}^2 + m_3^2 \tilde{h}_{ii}^2 - 
  2 m_4^2 \tilde{h}_{00}\tilde{h}_{ii} \right), 
\end{equation}
($0$ and $i=1,2,3$ are the time and space indices, respectively)
that can induce, for a suitable combination of the masses $m_i$'s 
\cite{rubakov}, a Yukawa term in the potential $\sim (1/r)e^{-r/r_m}$
in weak field limit approximation. 
 
From these solutions,
in general it follows that a test particle of ordinary matter (type 1)
feels a static potential induced by
a point-like source containing the mass fractions $\mathcal{M}_1$ and
$\mathcal{M}_2$ of the two types of matter, as 
\begin{equation} \label{pot01}
  \phi(r)= \frac{G}{2r}  
  \left(\mathcal{M}_1+\mathcal{M}_2\right)
  + \xi  \frac{G e^{-r/r_m}}{2r} 
  \left(\tan^2\vartheta  \mathcal{M}_1-\mathcal{M}_2 \right),
\end{equation}
where $G = 1/(8\pi(M_1^2 + M_2^2))$ and $\xi$ and $r_m$ are
parameters that depend on the pattern of the masses
in the (\ref{rub}). The first term is mediated
by massless gravity and the second term by the massive one. 
The symmetric case, i.e. $M_1=M_2$,
allow us to simplify the potential (\ref{pot01}) in the
following form
\begin{equation} \label{pot02}
  \phi(r) = \frac{G \mathcal{M}_1}{r} \left(\frac{1+\xi 
  e^{-r/r_m}}{2} \right) +
              \frac{G\mathcal{M}_2}{r} \left(\frac{1-\xi  
  e^{-r/r_m}}{2} \right),   
\end{equation}
that shows directly the modification with respect to
the Newtonian law (\ref{eq:cpot}).
Let us distinguish two cases: the mass term (\ref{rub}) is
Lorentz invariant or Lorentz breaking.  
In the first case the only consistent mass term (without ghosts)
is Pauli-Fierz type \cite{PF}
(i.e. $m_0=0$ and $m_{1,2,3,4}=m $).  
In this case in (\ref{pot01}) one has
$\xi=4/3$ and therefore a deviation  
from the Genral Relativity prediction for the light bending
in gravitational field
\footnote{
In this case one can define the discontinuity parameter:
$  \delta = 1 + (\sin^2 \vartheta)/3.$
Experimental limits on the post-Newtonian gravity \cite{will}
requires $\delta = 1.0000 \pm 0.0001$, 
therefore $\vartheta \simeq 10^{-2}$. As a consequence the 
Lorentz invariant case $M_1=M_2$ is excluded.},
also noted as van
Dan-Veltmann-Zakharov discontinuity (vDVZ).

In the second case, in general
the masses in (\ref{rub}) are different. 
The discontinuity is absent if, in the mass term, $m_0=0$,
$m_1\neq m_4$ and/or $m_2\neq m_3$. 
However, the bigravity (\ref{action}) 
produces a mass term 
with $m_1=0$  and therefore
does not give a Yukawa potential \cite{zurab}. 
The addition of a third auxiliary metric
$g_3$, which works as a bridge between $g_1$ and $g_2$ and decouples
with a Planck mass $M_3 \gg M_{1,2}$, allows us to have an effective
bigravity theory with $m_1\neq 0$ \cite{us01}. In this case 
the massive gravity introduces a Yukawa potential 
(with range $1/(\sqrt{3}m_4)$) 
without vDVZ discontinuity
(i.e. $\xi=1$) as in (\ref{bigr}). This potential,
in the mirror matter model, can explain the rotational curves
of galaxies with the condition $\mathcal{M}_2 \simeq 10 \mathcal{M}_1$ and
$r_m \simeq 20 $ kpc (i.e. $\Omega_D/\Omega_M \simeq 10$).  

\section{Rotational Curves of the Galaxies} 

The galaxy rotational curves
describe  the velocity of
stars and interstellar gases as a function  of the distance $ r $ from the 
center. For the sake of simplicity we apply our model to  disk
galaxies, where 
most of the matter (about 2/3) 
is concentrated in the inner region, called bulge. 
Indeed, one can 
suppose that the matter density along the 
profile follows approximately the luminosity, which decreases
exponentially with $ r $ moving out from the center. In a spherically 
symmetric approximation the visible matter has an exponential distribution 
depending only on $ r $, i.e. $\rho(r)
= \rho_0 e^{-r/r_0}$, where $\rho_0$ is the density in the central region
and $r_0$ is the size of the bulge. 
The Newtonian theory, that takes into account only visible baryonic
mass, does not explain the flattening of the rotational curves
at large distance and in general needs to introduce 
an extended halo composed by dark matter with different
density profile with respect to the visible one.

\begin{figure}
    \resizebox{0.60\columnwidth}{!}{\includegraphics{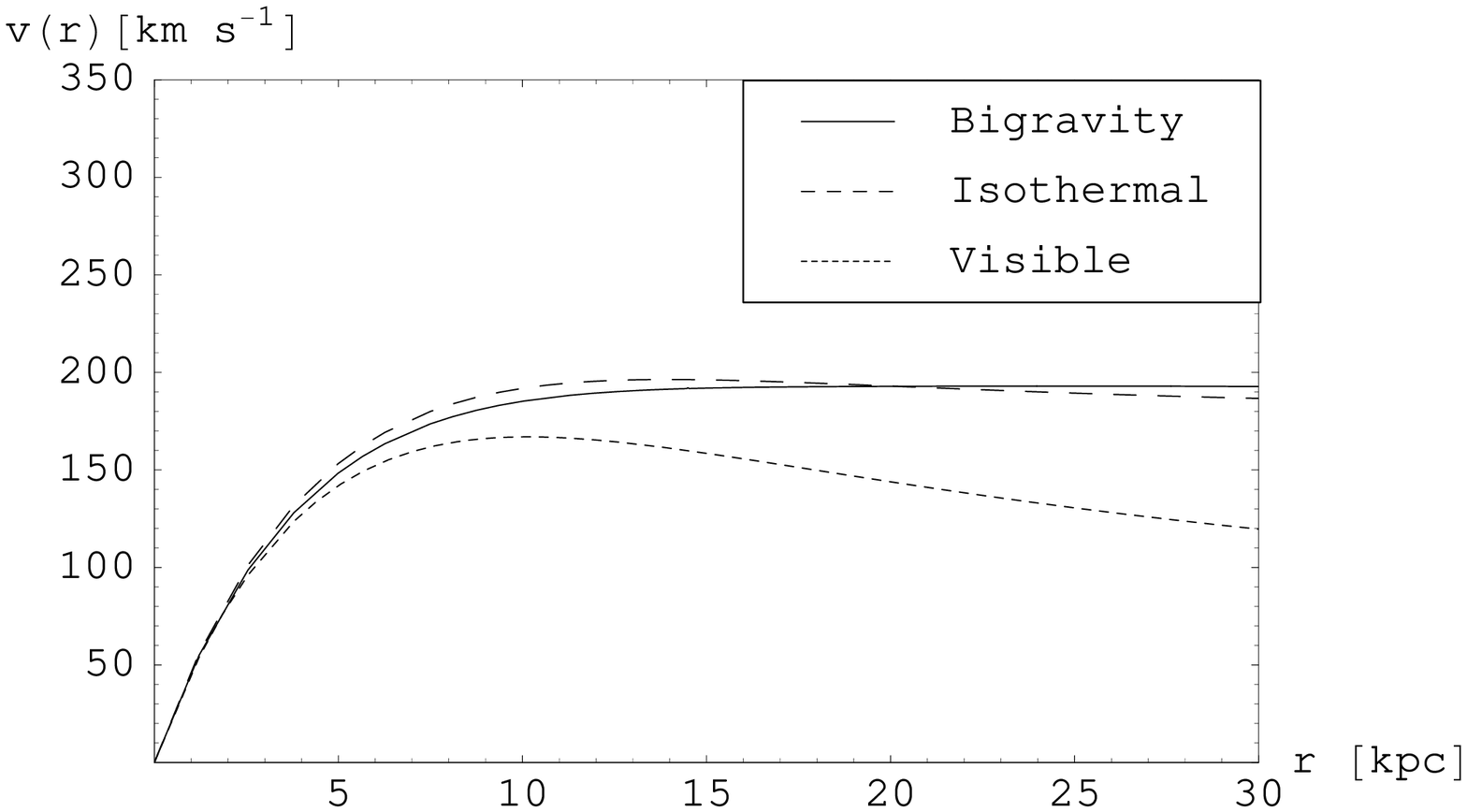}}
    \caption{Rotational curves fitted by the 
isothermal DM halos in standard gravity (dashed) 
and by the exponential DM profile in the bigravity
theory (solid) with 
$\mathcal{M}_2/\mathcal{M}_1=10$ and $r_m = 20 $. 
Contribution of the visible matter to the velocity given by 
the dotted curve.}
    \label{fig002}
\end{figure}

In bigravity theory, in which we consider the mirror matter as 
dark matter candidate, 
a different explanation emerges (\cite{us02}). 
Let us assume the following hypotheses
\begin{itemize}
  \item The ordinary and dark matters interact only via gravity,
        modified according to the bigravity model (\ref{bigr}).

  \item The two types of matter have similar density profiles in the galaxy,
        i.e. exponential along the disk, $\rho(r) \propto \exp (-r/r_0)$.
\end{itemize}
The velocity of an object at distance $ r $ is determined by equating
the centrifugal acceleration with the radial component of 
gravitational acceleration $a(r)$ 
derived from the potential (\ref{bigr}). For instance, from the 
gravitational field of point-like source of both types sitting in the 
center, we find
\begin{equation} \label{eq:big_for}
  a(r) = 
  G_N \left[ \frac{\mathcal{M}_1+\mathcal{M}_2}{2r^2} +
  \frac{\mathcal{M}_1-\mathcal{M}_2}{2r^2}
  \left(1+\frac{r}{r_m} \right) e^{-\frac{r}{r_m}} \right].  
\end{equation}
In order to obtain the total force on a
star moving approximately along a circular orbit around the galaxy center
we have to integrate (\ref{eq:big_for}) on the matter spatial
distribution.
Notice that for $ r << r_m $, the influence of matter of type 2 on a particle 
1 is negligible and the behavior is Newtonian. The behavior in the opposite 
limit $ r >> r_m $ is also essentially Newtonian, though the test particle 
feels the presence of the total mass $ M_1 + M_2 $ and  the effective 
Newton constant is $ G/2 $. 
In the region $ r \sim r_m $ there is a 
significant deviation from the Newtonian theory due to the presence 
of matter of type 2, resulting  in a enhancement of $ v $. 
This result avoids the cusp problem which is present in the context of 
the CDM paradigm and reproduces  
a isothermal-like shape 
($\mathcal{M}_{\rm D}/\mathcal{M}_{\rm B}\simeq 5$
and $\rho_{\rm DM}(r) \propto(1+(r/r_0)^2)^{-1}$).
For instance, let us consider a galaxy with $M_1=10^{11}M_{\odot}$ 
and the bulge size $r_0\simeq 3$ kpc (e.g. the Milky Way).
In Fig. \ref{fig002}
we compare the
rotational curves fitted with the 
isothermal DM halo ($a\simeq 8$ kpc)
in standard gravity and  
those fitted with the exponential DM profile in the bigravity
theory ($r_0=5.4$ kpc for the invisible distribution).
Notice that both curves have approximately the same behavior.
In addition, we show the visible matter contribution to the velocity
reproducing the Keplerian fall proportional to $1/\sqrt{r}$.

\begin{figure}
    \resizebox{0.60\columnwidth}{!}{\includegraphics{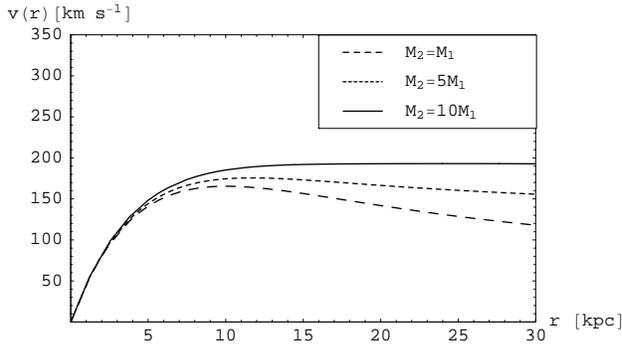}}
    \caption{The rotational curves for different
$\mathcal{M}_2/\mathcal{M}_1 $,
with $r_m = 20 $ kpc. Notice, that for 
$\mathcal{M}_1=\mathcal{M}_2 $ we obtain the Keplerian fall.}
    \label{fig001}
\end{figure}

The flat rotational curves can be reproduced varying the 
parameter $r_m$ and
the mass ratio $\mathcal{M}_2/\mathcal{M}_1$.
Fig. \ref{fig001} shows 
the rotational curves for different
$\mathcal{M}_2/\mathcal{M}_1 $,
with $r_m = 20 $ kpc. Notice, that for 
$\mathcal{M}_1=\mathcal{M}_2 $ we obtain the Keplerian fall
as we expect from the Newtonian potential (\ref{bigr}). 
Incidentally, if both the sectors have 
the same content of matter the potential (\ref{bigr}) is indistinguishable
from the standard Newton potential $\phi=G(M_1/r)$ generated
by the visible matter.

\section{Conclusions}

If the Universe is made of two separate sectors, one visible and
one hidden, each of them can have
its own gravity and the two metrics can interact
as in a bigravity pattern. This lead to a large distance
modification of the gravitational force, though it remains
Newtonian at small distances where each type of matter 
feels only itself, in agreement with the precision test
of General Relativity in the Solar System.

The mirror matter as dark matter, together with Lorentz breaking bigravity
can explain the flattening of rotational curves of galaxy  
at large distance. This model supposes that both dark matter
and modification of gravity are present. The main advantage
of this theory is that, due to the same property of both types of 
matter, the dark matter can have a mass distribution in galaxy 
similar to the visible one. This avoids the need to invoke the 
presence of extended halo distributions that are in conflict
with numerical simulations in the CMD paradigm.
 
As an additional result, it interesting to note that 
the Newton constants for type~1~-~type~1 attraction differs by a factor 2
for large and small distances:
$G_N(r \ll r_m)=G $ and  $G_N(r \gg r_m)=G/2 $.
In the general potential (\ref{pot01}) the Newton constant can be measured as 
$G_N(r \ll r_m)=G[(1+\xi \tan^2\vartheta)/2]$ between type~1~-~type~1 matter
at small distance, while at large distance we have $G_N(r \gg r_m)=G$. 
Moreover, it is interesting to note that,
if $\xi > 1$, at small distance $r \ll r_m$ one has antigravity
between type 1 and type 2 objects.

\section*{Acknowledgments}

I thank the organizers of the International Conference 
ACFC'2007. 
I thank Z. Berezhiani, F. Nesti and L.Pilo for collaboration on
refs. \cite{us01,us02} and P. Salucci for interesting remarks.
This work is partially supported by the MIUR grant under the Project 
of National Interest PRIN 2006 ``Astroparticle Physics'' and by
the European FP6
Network "UniverseNet" MRTN-CT-2006-035863.

\end{document}